\documentclass[useAMS,usenatbib]{mn2e}

\usepackage {graphicx}
\usepackage {times}

\title[AKARI-SDSS IR LFs]
{Infrared Luminosity Functions of AKARI-SDSS  Galaxies
}
\author[Goto et al.]{Tomotsugu Goto$^{1,2}$, 
 \thanks{E-mail:tomo@ifa.hawaii.edu} 
Stephane Arnouts$^3$,
Matthew Malkan$^{4}$,
Toshinobu Takagi$^{5}$,
Hanae Inami$^{6}$,
\newauthor 
Chris Pearson$^{7}$,
Takehiko Wada$^{5}$,
Hideo Matsuhara$^{5}$,
Chisato Yamauchi$^{5}$,
\newauthor 
Tsutomu T. Takeuchi$^{10,11}$,
Takao Nakagawa$^{5}$,
Shinki Oyabu$^{11}$,
Daisuke Ishihara$^{11}$,
\newauthor 
David B. Sanders$^{1}$,
Emeric Le Floc'h$^{12}$,
Hyung Mok Lee$^{13}$,
Woong-Seob Jeong$^{14}$,
\newauthor 
Stephen Serjeant$^{15}$,
and
Chris Sedgwick$^{15}$
\\
$^{1}$Institute for Astronomy, University of Hawaii, 2680 Woodlawn Drive, Honolulu, HI, 96822, USA\\
$^{2}$Subaru Telescope 650 North A'ohoku Place Hilo, HI 96720, USA\\
$^{3}$Canada France Hawaii Telescope, 65-1238 Mamalahoa Hwy, Kamuela, Hawaii 96743 USA \\
$^{4}$Department of Physics and Astronomy, UCLA, Los Angeles, CA, 90095-1547, USA\\
$^{5}$Institute of Space and Astronautical Science, Japan Aerospace Exploration Agency, 	     Sagamihara, Kanagawa 252-5210\\
$^{6}$Spitzer Science Center, California Institute of Technology, Pasadena, CA 91125, USA\\
$^{7}$ RAL Space, Rutherford Appleton Laboratory, Chilton, Didcot, Oxfordshire OX11 0QX, UK\\
$^{8}$ Institute for Space Imaging Science, University of Lethbridge,Lethbridge, Alberta T1K 3M4, Canada\\
$^{9}$ Astrophysics Group, Department of Physics, The Open University, Milton Keynes, MK7 6AA, UK\\
$^{10}$Institute for Advanced Research, Nagoya University, Furo-cho, Chikusa-ku, Nagoya 464-8601\\
$^{11}$Division of Particle and Astrophysical Science, Nagoya University, Furo-cho, Chikusa-ku, Nagoya 464-8602, Japan \\
$^{12}$CEA-Saclay, Service d'Astrophysique, France\\
$^{13}$Department of Physics \& Astronomy, FPRD, Seoul National University, Shillim-Dong, Kwanak-Gu, Seoul 151-742, Korea	\\
$^{14}$Korea Astronomy and Space Science Institute 61-1, Hwaam-dong, Yuseong-gu, Daejeon, Republic of Korea 305-348\\
$^{15}$Astrophysics Group, Department of Physics,  The Open University, Milton Keynes, MK7 6AA, UK
}
\begin{document}
\maketitle
\label{firstpage}
\begin{abstract}
 By cross-correlating AKARI all sky survey in 6 infrared (IR) bands (9, 18, 65, 90, 140, and 160$\mu m$) with the SDSS galaxies, we identified 2357 infrared galaxies with a spectroscopic redshift.
 This is not just one of the largest samples of local IR galaxies, but  AKARI provides crucial FIR bands in accurately measuring galaxy SED across the peak of the dust emission at $>100\mu m$.
  By fitting modern IR SED models to the AKARI photometry, we measured the total infrared luminosity ($L_{IR}$) of individual galaxies. 

 Using this $L_{IR}$, we constructed the luminosity functions of infrared galaxies at a median redshift of z=0.031. 
 The LF agrees well with that at z=0.0082 (the RBGS),
showing smooth and continuous evolution toward higher redshift LFs measured in the AKARI NEP deep field.
 By integrating the IR LF weighted by $L_{IR}$, we measured the local cosmic IR luminosity density of 
$\Omega_{IR}$= (3.8$^{+5.8}_{-1.2})\times 10^{8}$ $L_{\odot}$Mpc$^{-3}$.

We separate galaxies into AGN (active galactic nuclei), star-forming, and composite by using the  $[NII]/H\alpha$ vs $[OIII]/H\beta$ line ratios.
  The fraction of AGN shows a continuous increase with increasing $L_{IR}$ from 25\% to 90\% at 9$<log L_{IR}<$12.5.
 The SFR$_{H\alpha}$ and $L_{[OIII]}$ show good correlations with $L_{IR}$ for SFG (star-forming galaxies) and AGN, respectively.
 The self-absorption corrected H${\alpha}$/H${\beta}$ ratio shows a weak increase with   $L_{IR}$ with a substantial scatter.
 When we separate IR LFs into contributions from AGN and star-forming galaxies (SFG), the AGN contribution becomes dominant at $L_{IR}>10^{11}L_{\odot}$,  coinciding the break of the both SFG and AGN IR LFs. At $L_{IR}\leq10^{11}L_{\odot}$, SFG dominates IR LFs.
 Only 1.1$\pm0.1$\% of $\Omega_{IR}$ is produced by LIRG ($L_{IR}>10^{11}L_{\odot}$), and 
 only 0.03$\pm$0.01\% is by  ULIRG ($L_{IR}>10^{12}L_{\odot}$) in the local Universe. 
 Compared with high redshift results from the AKARI NEP deep survey, we observed a strong evolution of 
$\Omega_{IR}^{SFG}\propto$(1+z)$^{4.1\pm0.4}$ and  
$\Omega_{IR}^{AGN}\propto$(1+z)$^{4.1\pm0.5}$. 
Our results show all of our measured quantities (IR LFs, $L^*$,  $\Omega_{IR}^{AGN}$,  $\Omega_{IR}^{SFG}$) show smooth and steady increase from lower redshift  (the RBGS) to higher redshift (the AKARI NEP deep survey).
\end{abstract}


\section{Introduction}

\begin{figure}
\begin{center}
\includegraphics[scale=0.6]{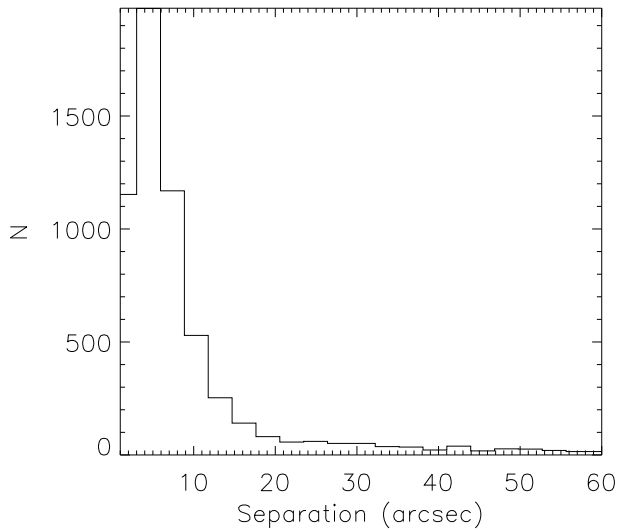}
\end{center}
\caption{
 Angular separation between the AKARI FIS and SDSS positions. 
 We use galaxies matched within 20 arcsec for LFs.
}\label{fig:sep}
\end{figure}

To understand the cosmic history of star formation and AGN  (active galactic nuclei), it is vital to understand infrared (IR) emission; the more intense star formation, the more deeply it is embedded in the dust, hence, such star formation is not visible in UV but in the infrared. Similarly, AGN evolutionary scenarios predict that  AGN are heavily obscured at their youngest, compton-thick stage \citep{2009ApJ...696..110T}. 
 The Spitzer and AKARI satellites revealed a great deal of infrared emission in the high-redshift Universe, showing strong evolution in the infrared luminosity density \citep{2005ApJ...632..169L,2005ApJ...630...82P,2006MNRAS.370.1159B,2007ApJ...660...97C,2009A&A...496...57M,2010A&A...518L..27G,2010A&A...515A...8R}. 
 For example, at z=1, \citet{Goto_NEP_LF} estimated 90\% of star formation activity is hidden by dust.

However, to investigate evolution of IR LF, these high-redshift studies need a good comparison sample at z=0.
Even today, the often used is the IRAS LFs \citep{1993ApJS...89..1R,2003AJ....126.1607S,Goto_RBGS_LF} from 1980s, with only several hundred galaxies.
 In addition, for more than 25 years, bolometric infrared luminosities ($L_{IR,8-1000\mu m}$) of local galaxies have been estimated using equation in  \citet{1987PhDT.......115P}, which is a simple polynomial, obtained assuming a simple blackbody and dust emissivity. Furthermore, the reddest filter of IRAS is 100$\mu m$, which does not span the peak of the dust emission for most galaxies, leaving a great deal of uncertainty. 
A number of studies found cold dust that cannot be detected with the IRAS. For example, \citet{2001MNRAS.327..697D} detected such cold dust with $T\sim20$K using Scuba 450,850 $\mu$m flux. 
\citet{2009MNRAS.397.1728S} detected cold galaxies with SED peaks at longer wavelengths using Spitzer/MIPS. These results cast further doubt on  $L_{IR}$ estimated using only $<100\mu$m photometry.  
More precise estimate of local $L_{IR}$ and thus the local IR luminosity function (LF) have been long awaited, to be better compared with high redshift work and to understand where the end-point of the cosmic IR density evolution is.

AKARI, the Japanese infrared satellite \citep{2007PASJ...59S.369M}, provides the first chance to  rectify the situation since IRAS; AKARI performed an all-sky survey in two mid-infrared bands (centered on 9 and 18 $\mu m$) and in four far-infrared bands (65,90, 140, and 160$\mu m$). Its 140 and 160$\mu m$ sensitivities are especially important to cover across the peak of the dust emission, allowing us to accurately measure the Rayleigh-Jeans tail of the IR emission. 

Using deeper data and modern models, in this work, we aim to measure local $L_{IR}$, and thereby the IR LF.
By matching the AKARI IR sources with the SDSS galaxies, our sample contains $\sim$2357 IR galaxies. 
Compared to the previous work  \citep{2003AJ....126.1607S,Goto_RBGS_LF}, the sample is several times larger, allowing us to accurately measure IR LF of the local Universe.
 The optical spectra of the SDSS also allows us to separate IR emission from AGN and star-forming galaxies.
 This work provides us an important local benchmark to base our evolution studies at high redshift both by the current AKARI, Spitzer, and Herschel satellites, and by next-generation IR satellites such as WISE, JWST, and SPICA.
Following \citet{2003AJ....126.1607S}, we adopt a cosmology with $(h,\Omega_m,\Omega_\Lambda) = (0.75,0.3,0.7)$.

\section{Data and Analysis}\label{Data}
\subsection{AKARI-SDSS sample}

\begin{figure}
\begin{center}
\includegraphics[scale=0.4]{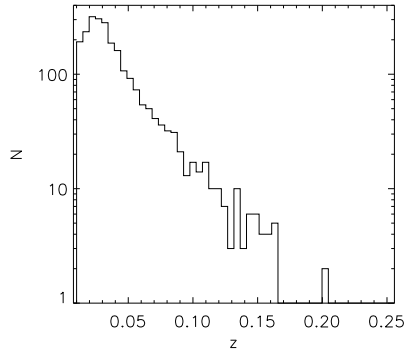}
\end{center}
\caption{Redshift distribution of the AKARI-SDSS galaxies.
}\label{fig:zhist}
\end{figure}

 AKARI \citep{2007PASJ...59S.369M} performed an all-sky survey in two mid-infrared bands (centered on 9 and 18 $\mu m$) and in four far-infrared bands (65,90,140, and 160$\mu m$). 
In this work, we use the  version 1 of the AKARI/IRC point source catalog and the AKARI/FIS bright source catalog, which is selected in 90$\mu$m.
 The 5 $\sigma$ sensitivities in the AKARI IR filters ($S9W,L18W,N60,WS,WL$ and $N160$) are 0.05, 0.09, 2.4, 0.55, 1.4, and 6.3 Jy \citep{IRCpaper,2009AIPC.1158..169Y}.
 In addition to the much improved sensitivity and spatial resolution over its precursor (the IRAS all-sky survey), the presence of 140 and 160$\mu m$ bands is crucial to measure the peak of the dust emission in the FIR wavelength, and thus the $L_{IR}$ of galaxies.

To measure $L_{IR}$, we need spectroscopic redshift of individual galaxies. 
We have cross-correlated the AKARI FIS bright source catalog with the SDSS DR7 galaxy catalog \citep{2009ApJS..182..543A}, which is the largerst redshift survey in the local Universe to date.

 We used a matching radius of 20 arcsec, as shown in Fig.\ref{fig:sep}.
This radius is determined based mainly on the positional accuracy of the AKARI FIR sources since for the SDSS, positional accuracy is less than 0.1'' to the survey limit \citep[$r\sim$22][]{2003AJ....125.1559P}. 
 AKARI's PSF size in the 90$\mu$m filter is 39$\pm$1''. 
 The astrometric accuracy of the AKARI FIS bright source catalog is 3.8'' in cross-scan direction and 4.8'' in in-scan direction \citep{2009AIPC.1158..169Y}. 
 The 20'' radius also accounts for the physiscal shift that could be present between the peaks of the FIR and the optical emission.
 At the median redshift of z=0.031, 20'' corresponds to 12 kpc.
  The source density of the SDSS main spectroscopic galaxies is 92 deg$^{-2}$ \citep{2002AJ....124.1810S}.  
 Therefore, within the 20'' around each AKARI sources, 0.009 of background galaxies are expected to be present, i.e., less than 1\% of our sources are expected to be a chance coincidence, which should not affect our estimate of the LFs. See \citep{1992MNRAS.259..413S} for a maximum likelihood approach to this.


 We do not use galaxies at $z<0.01$ since (i) at such a low velocity, uncertainty from peculiar velocity is too large to estimate distance from spectroscopic redshift, and (ii) the AKARI's photometry of largely extended objects is not finalized yet. 
 We also remove galaxies at $z\geq0.3$ to avoid too much evolutionary effect within the redshift range.
 We also use SDSS galaxies with extinction corrected $r_{petro}\leq 17.7$, where the SDSS main galaxy spectroscopic target selection can be considered complete. 
 To obtain reliable LFs, we only use AKARI sources with $S_{90\mu m}>0.7 Jy$, where completeness is greater than 80\%.
Among these, 99\% of sources have high quality flag of {\ttfamily fqual=3} in  $90\mu m$. 
 67\% of AKARI sources with  $S_{90\mu m}>0.7 Jy$ in the area covered by the SDSS had an optical counterpart. The rest of the sources are either stars or galaxies fainter than $r_{petro}=17.7$, which are property accounted for the LF computation in Section \ref{sec:vmax}.

 Previous sample of local IR LFs was based on IRAS $S_{60\mu m}$ flux, and thus, it has been a concern that colder galaxies, whose FIR emission peak at larger wavelengths, might have been missed in the sample. 
 Our sample is selected using  $S_{90\mu m}$ flux, and thus, more robust to such a selection bias.

 We found cross-matches of 2357 
 galaxies, with which we will measure IR LFs. 
Among these,  97, 98, and 87\% of sources have measured flux in 65, 140 and 160$\mu$m.
 For these source the AKARI IRC catalog is also cross-matched. Only 0.8\% of FIS-SDSS sources do not have an IRC counterpart within 20 arcsec.
 Fig.\ref{fig:zhist} shows a redshift distribution of the sample.
 The number of galaxies used are similar to the IRAS based work by \citet{2005MNRAS.360..322G}, despite the twice increase in the optical (SDSS) sample. This is due to more stringent cross-matching criterion we used. AKARI's improved astrometry allowed us to remove ambiguous cross-ids.
 Nevertheless, thanks to the large sky coverage of the SDSS and AKARI surveys, this is one of the largest number of galaxies used to construct IR LFs. For example, previous local IR LFs are based on $\sim$600 galaxies \citep{2003AJ....126.1607S,Goto_RBGS_LF} .


\subsection{Estimating total IR luminosity}\label{sec:LIR}

\begin{figure}
\begin{center}
\includegraphics[scale=0.6]{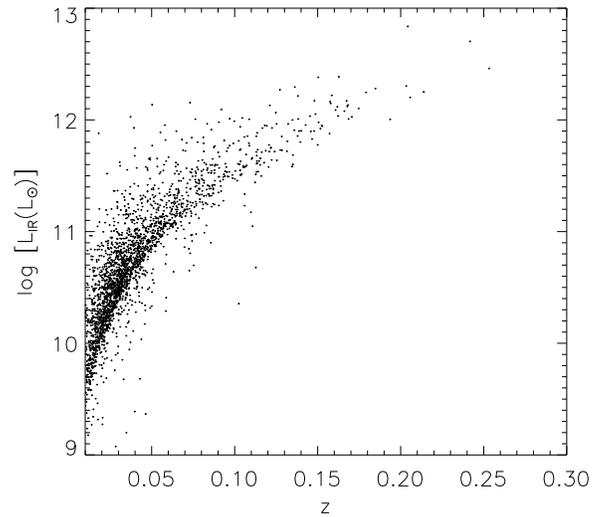}
\end{center}
\caption{
 $L_{IR}$ measured by the AKARI as a function of spectroscopic redshift.
}\label{fig:z_vs_Ltir}
\end{figure}


For these galaxies, we estimated total IR luminosities ($L_{IR}$) by fitting the AKARI photometry with SED templates. 
We used the {\ttfamily  LePhare} code\footnote{http://www.cfht.hawaii.edu/$^{\sim}$arnouts/lephare.html} to fit the infrared part ($>$7$\mu$m) of the SED. 
We fit our AKARI FIR photometry with the SED templates from \citet[CHEL hereafter][]{2001ApJ...556..562C}, which showed most promising results among SED models tested by \citet{Goto_RBGS_LF}. It is a concern that the CHEL models do not include SEDs for AGN. However, we show later in Section 2.3 that this is not a major problem using observed data. 
We did not include IRAS photometry in the SED fit partly because the better spatial resolution of AKARI often resolves background cirrus and nearby companion better \citep{2007PASJ...59S.429J}, and partly because not all of the AKARI sources are detected with the IRAS due to the difference in flux limits.
  Although the shape of these SEDs are luminosity-dependent, the large baseline from AKARI observations  ($S9W,L18W,N60,WS,WL$ and $N160$) allows us to  adopt a free scaling to obtain the best SED fit,  which is then rescaled to derive $L_{IR}$.  
To be precise, we only used $\geq 65\mu$m flux to free scale the SEDs since mid-IR bands can sometimes be affected by the stellar emission.
For those sources with low quality flag ({\ttfamily fqual=1}), we adopted a minimum error of 25\%.
 In this work,  $L_{IR}$ is measured in the wavelength range of 8-1000$\mu$m.
Fig. \ref{fig:z_vs_Ltir} shows  $L_{IR}$ as a function of spectroscopic redshift.

%
%

\subsection{ AGN/SFG separation}\label{sec:f_agn}

\begin{figure}
\begin{center}
\includegraphics[scale=0.6]{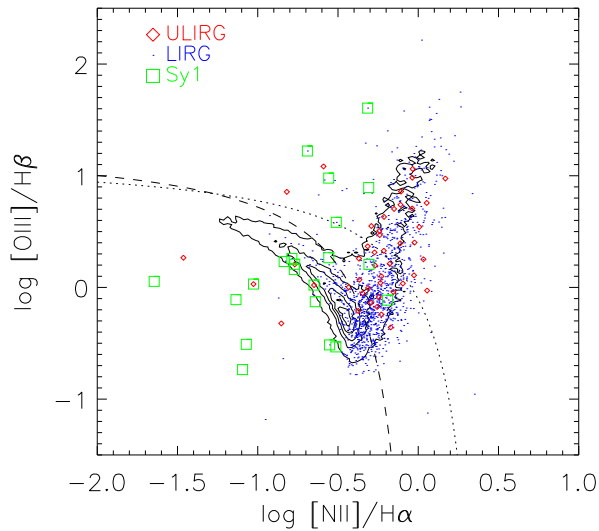}
\end{center}
\caption{
Emission line ratios used to select AGNs from our sample. The contour shows distribution of all galaxies in the SDSS with $r<17.77$ (regardless of IR detection). 
The dotted line is the criterion between starbursts and AGNs described in \citet{2001ApJ...556..121K}.
The dashed line is the criterion by \citet{2003MNRAS.346.1055K}. 
Galaxies with line ratios higher than the dotted line are regarded as AGNs. 
Galaxies below the dashed line are regarded as star-forming. 
Galaxies between the dashed and dotted lines are regarded as composites. 
The blue and red dots are for ULIRGs, LIRGs, respectively.
The green squares are Seyfert 1 galaxies identified by eye-balling optical spectra.
}\label{fig:BPT}
\end{figure}

One has to be careful that two different physical mechanisms contribute to total infrared emission of galaxies: one by star-formation and the other by AGN. It is therefore fundamental to separate IR contribution from these two different physics, to understand the evolution of the star-formation and AGN activity. However, this has been notoriously difficult and has been a subject of active debate over the last decades. Many AGN/SF separation methods have been proposed such as X-ray, radio luminosity, optical line ratios, PAH strengths, submm properties and so on \citep{2008ApJ...686..155Z,2008ApJ...683..659S,2010ApJ...720..786L}. Often, not all of the indicators do not agree one another,  partly due to complicated nature of IR sources (multiple core, composite ...etc).

 

In this work, since we have optical spectra of individual galaxies, 
we use $[NII]/H\alpha$ vs $[OIII]/H\beta$ line ratios to classify galaxies into AGN or SFG (star-forming galaxies). 
In Fig.\ref{fig:BPT}, we plot  $[NII]/H\alpha$ against $[OIII]/H\beta$ line ratios. 
The black contours show the distribution of all emission line galaxies in the SDSS, regardless of the AKARI detection.
The dotted and dashed lines are the AGN/SFG separation criteria presented by \citet{2001ApJ...556..121K} and \citet{2003MNRAS.346.1055K}, respectively.
We regard galaxies above the dotted line as AGN, below the dashed line as SFG, and those in between as composite galaxies.

Although these line diagnosis criteria work well for narrow-line AGN, they could be problematic for broad-line AGN. Although, in principle, one could separate narrow line components from broad-line by multiple-line fitting, it is often difficult with weak lines on noisy spectra. 
 To see how well broad line AGN are classified in Fig.\ref{fig:BPT}, 
 one of us (M.M.) has eyeballed all emission line IR galaxies with FWHM(H$\alpha$) greater than FWHM([OIII]) by more than 150 km/s.
 We identified 22 broad-line AGN, which are shown in green squares in Fig.\ref{fig:BPT}.
 As expected, the broad-line AGN are scattered everywhere in the Figure and escapes from the above criteria.
 Therefore, we manually overruled the classification of these broad-line AGN, and treated them as AGN in the rest of the paper, regardless of the line ratios in Fig.\ref{fig:BPT}.
 In this process, we also came across 20 narrow line AGN (although these are not a complete sample). They were successfully classified as AGN in Fig.\ref{fig:BPT}.

\begin{figure}
\begin{center}
\includegraphics[scale=0.6]{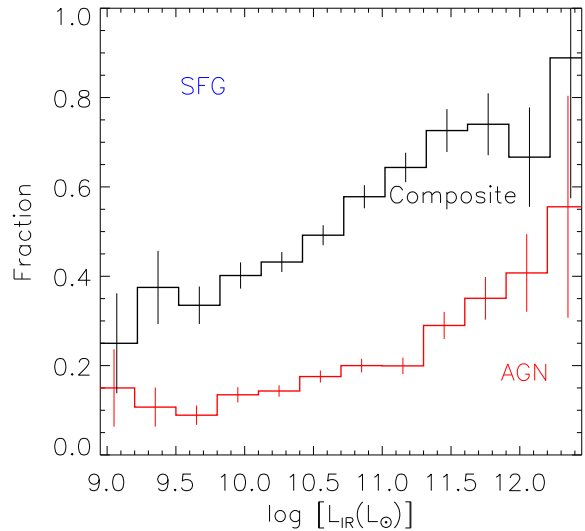}
\end{center}
\caption{Fractions of AGN and composite galaxies as  a function of $L_{IR}$.
AGN are classified using \citet{2001ApJ...556..121K}    among galaxies with all 4 lines measured.  Composite galaxies include those classified as AGN using \citet{2003MNRAS.346.1055K}.
}\label{fig:AGN_fractions}
\end{figure}

In Fig.\ref{fig:BPT}, ULIRGs and LIRGs are marked with red diamonds and blue dots, respectively.
It is interesting that majority of (U)LIRGs are aligned along the AGN branch of the diagram, implying the 
AGN fraction is high among (U)LIRGs. This is more clearly seen in Fig.\ref{fig:AGN_fractions}, where we plot fractions of AGN as a function of $L_{IR}$.
The red solid line shows fractions of AGN only, and the black dotted line includes composite galaxies.
Note that the fractions do not include completely obscured sources in optical for both SFG and AGN.
The fractions of AGN (including composites) increase from 15\% (25\%) at log$L_{IR}$=9.0 to  55\% (90\%) at  log$L_{IR}$=12.3.
 This results agree with previous AGN fraction estimates \citep{2005MNRAS.360..322G,2010ApJ...709..884Y,2010ApJ...709..572K}.
 Improvement in this work is that due to much larger statistics, we were able to show fractions of AGN in much finer luminosity bins, more accurately quantifying the increase. Especially, a sudden increase of $f_{AGN}$ at log$L_{IR}>$11.3 is notable due to the increased statistics in this work.


\begin{figure}
\begin{center}
\includegraphics[scale=0.6]{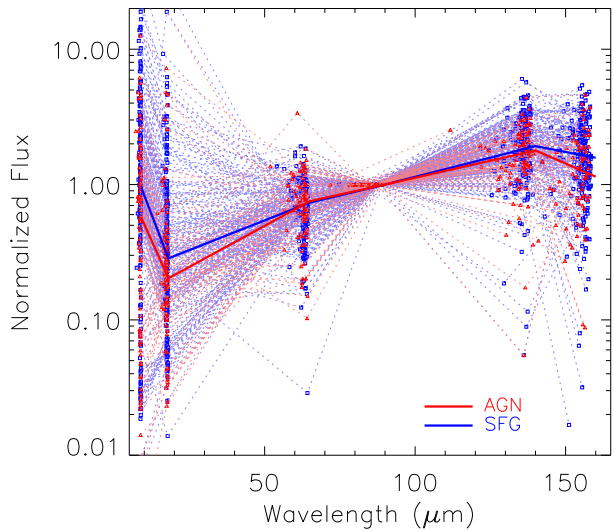}
\end{center}
\caption{SEDs of AGN (red triangles) and SFG (blue squares).
The red and blue solid lines connect median values for AGN and SFG, respectively.
Note that only galaxies detected in all the 6 AKARI bands are shown here.
The flux are normalized at 90$\mu$m.
}\label{fig:SED}
\end{figure}

 Having AGN/SFG classified, in Fig.\ref{fig:SED}, we show restframe SEDs of AGN (red triangles) and SFG (blue squares), separately. 
 Note that only galaxies detected in all the 6 AKARI bands are shown here. 
The flux are normalized at 90$\mu$m. The solid lines connect median points for each sample.
 In Fig.\ref{fig:SED}, there is no significant difference between AGN and SFG SEDs. This has two implications: although the CHEL SED models do not include SEDs for AGNs, the SED fit can be performed using the galaxy SED models. At the same time, we need to be careful that although we selected AGN based on the optical line ratios, infrared SEDs of these optically-selected AGN can be dominated by star-formation activity in the AGN host galaxy.

\begin{figure}
\begin{center}
\includegraphics[scale=0.4]{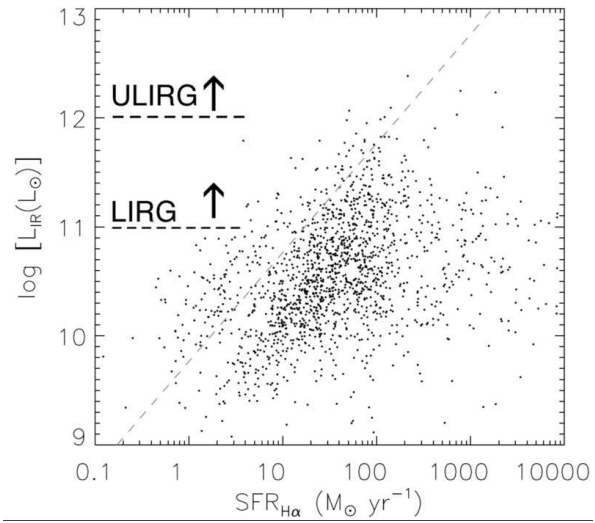}
\end{center}
\caption{
The SFR computed from the H$\alpha$ emission line is plotted against the total infrared luminosity ($L_{IR}$). 
The H$\alpha$ lines are corrected for stellar absorption and extinction using the Balmer decrement.
AGNs are excluded from this figure using Fig.\ref{fig:BPT}. 
}\label{fig:SFR_ha}
\end{figure}

In Fig.\ref{fig:SFR_ha}, we show star formation rate (SFR) computed from the self-absorption and extinction corrected $H{\alpha}$ flux \citep{2005MNRAS.360..322G} against $L_{IR}$ for SFGs. The SFR is also corrected for the fiber loss of the SDSS spectrograph.
There is a correlation, but compared with the well-known relation by \citet{1998ARA&A..36..189K}, there is a significant offset.
 This might warn us a simple application of \citet{1998ARA&A..36..189K}'s law to IR luminous galaxies. \citet{1998ARA&A..36..189K}'s sample was dominated by more regular (less IR luminous) spiral galaxies, whose $L_{IR}$ might stem from cool cirrus clouds warmed by older stars. This might not be the case for vigorously star-forming LIRG and ULIRGs with hot dust.

\begin{figure}
\begin{center}
\includegraphics[scale=0.6]{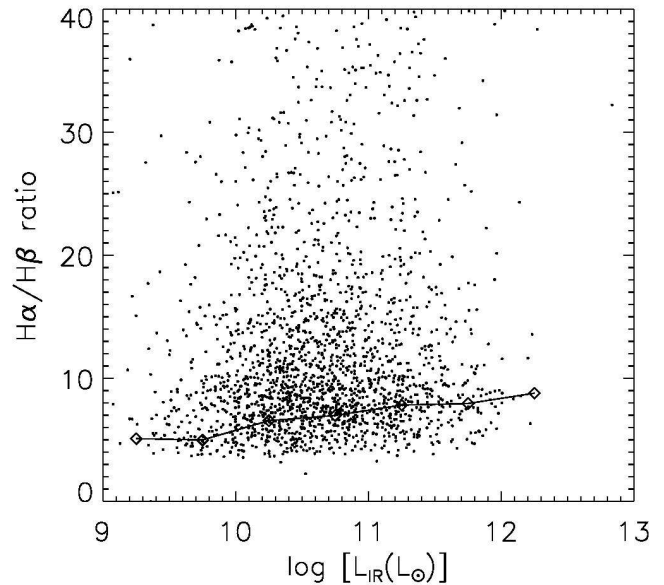}
\end{center}
\caption{The self-absorption corrected H$\alpha$/H$\beta$ ratios are shown against $L_{IR}$ for SFG.  The solid line connects median values.
}\label{fig:Hab}
\end{figure}

In Fig.\ref{fig:Hab}, we plot  the self-absorption corrected $H\alpha$/ $H\beta$ ratio as a function of $L_{IR}$ for SFG.
Although the median in the solid line shows a slight increase with increasing  $H\alpha$/ $H\beta$ ratio, but with a large scatter, suggesting that not only $L_{IR}$ decides the dust extinction in galaxies.

\begin{figure}
\begin{center}
\includegraphics[scale=0.6]{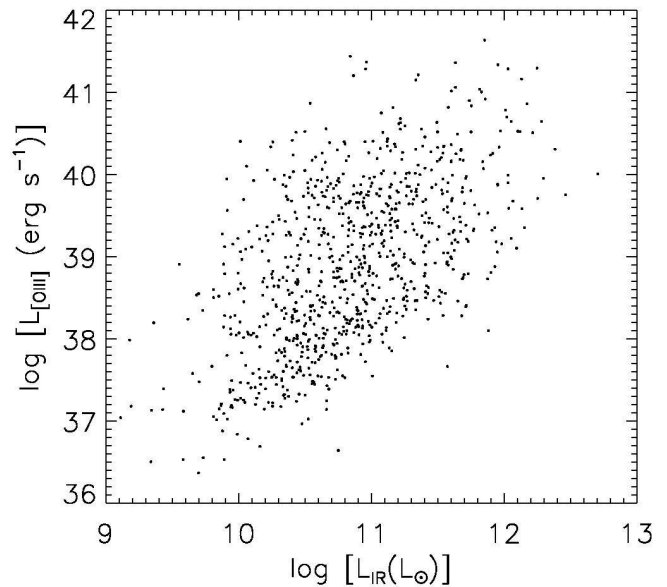}
\end{center}
\caption{  Luminosity of the [OIII](5008\AA) emission line is plotted against total infrared luminosity for AGNs.
Only objects classified as AGN are shown here.}\label{fig:oiii_lir} 
\end{figure}

In Fig.\ref{fig:oiii_lir} we plot   $L_{IR}$ vs $L_{[OIII]}$. Only objects classified as AGN are shown here. Both of these luminosities are considered to be good indicators of the AGN power, and 
there is a recognizable correlation, however, with a significant scatter. The exact reason of the scatter is unknown but possible causes include contamination from HII regions, cirrus emission, shocks, and  emission from very little narrow-line region directly powered by an accreting black hole.

\begin{figure}
\begin{center}
\includegraphics[scale=0.6]{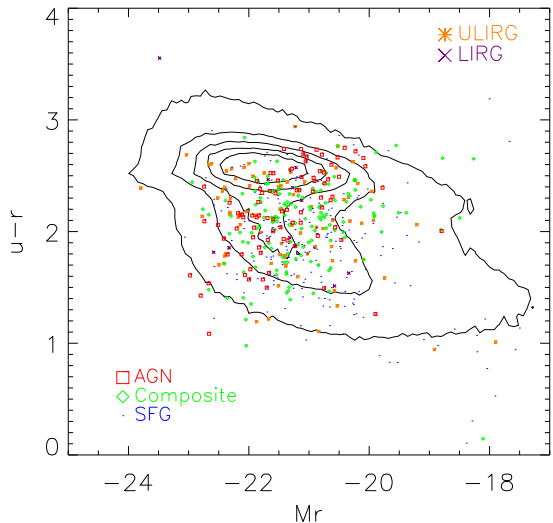}
\end{center}
\caption{Restframe $u-r$ vs $Mr$ color-magnitude diagram. The contours show the distribution of all SDSS galaxies with $r<17.7$.
  The orange stars and crosses denote ULIRGs and LIRGs in our sample.
 The red square, green diamond, and blue dots are for AGN, Composite, and SF IR galaxies classified using Fig.\ref{fig:BPT}.
}\label{fig:cmd}
\end{figure}

In Fig.\ref{fig:cmd}, we show restframe $u-r$ vs $M_r$ color-magnitude diagram. We used \citet[][v4\_1]{2003AJ....125.2348B} for k-correction.
 The contours show the distribution of all SDSS galaxies with $r<17.7$ regardless of IR detection. 
ULIRGs and LIRGs do not follow neither the red-sequence at $u-r\sim2.6$ nor blue cloud at $u-r<2.0$, but widely distributed in the $u-r$ color.
As expected, ULIRGs are more luminous in $M_r$ than LIRGs.

 The red squares, green diamonds, and blue dots are for AGN, composite, and SF IR galaxies classified using Fig.\ref{fig:BPT}.
These IR galaxies also show wide distribution in Fig.\ref{fig:cmd}, with AGN slightly brighter than the composite, reflecting the fact that the AGN fractions are higher for more luminous objects (Fig.\ref{fig:AGN_fractions}). The SFG are slightly more extended to bluer clouds than AGN and the composite.
 These distributions are consistent with AGN and IR galaxies populate the so-called ``green valley'' between red-sequence and blue clouds \citep{2009ApJ...699L..43S,2010ApJ...711..284S}.
  Perhaps physical reasons deciding $u-r$ color are different for AGNs and SFG. For example, dust-reddening might play a significant role for SFG, especially infrared luminous ones. For AGN, lack of gas in red-sequence galaxies, and dominance of HII regions in blue-cloud galaxies may push AGN into the green valley.

We caution readers that our AGN/SFG classification is by no means complete, but merely one attempt. 
In reality most of galaxies have both AGN and star-formation, and thus contribution to $L_{IR}$ and $L_{[OIII]}$ from both components.
 When one is dominant, it is difficult to assess a contribution from the other, which inevitably becomes a contamination to the former.
Recently, it has been reported that to fit FIR SED of AGN dominant galaxies, one always need a starburst component, even for quasars \citep{2010A&A...518L..26S,2010A&A...518L..33H,2010A&A...517A..11P}.
 In this sense, our $L_{IR}$ of AGN include IR emission from an AGN host galaxy. The latter might be dominant in FIR wavelength.
 Those galaxies with line ratios consistent to be LINERs are treated as AGN in this paper, but their emission may have nothing to do with an accreting black hole, instead may be from shocks or something else. Therefore, it is possible that these classified as AGN in this paper still have significant infrared emission from host galaxies, instead of central black holes. To make the matter worse, it is additional unknown how many completely obscured galaxies we miss in our optical classification for both AGN and SFG. While this work had only optical line ratios for classification, the final conclusion must be based on multi-wavelength data.

\section{Infrared Luminosity functions}
\subsection{The 1/$V_{\max}$ method}\label{sec:vmax}

With accurately measured $L_{IR}$, we are ready to construct IR LFs.
Since our sample is flux-limited at $r=17.7$ and $S_{90\mu m}=0.7Jy$, we need to correct for a volume effect to compute LFs.
We used the 1/$V_{\max}$ method \citep{1968ApJ...151..393S}. An advantage of the 1/$V_{\max}$  method is that it allows us to compute a LF directly from data, with no parameter dependence or a model assumption. A drawback is that it assumes a homogeneous galaxy distribution and thus is vulnerable to local over-/under-densities \citep{2000ApJS..129....1T}. 

A comoving volume associated to any source of a given luminosity is defined as $V_{\max}=V_{z_{\max}}-V_{z_{\min}}$, where $z_{\min}$ is the lower limit of the redshift  and $z_{\max}$ is the maximum redshift at which the object could be seen given the flux limit of the survey. 
 In this work, we set $z_{\min}$=0.01 since at a very small redshift, an error in redshift measurement is dominated by a peculiar motion, and thus,  $L_{IR}$ also has a large error. 

For the infrared detection limit, we used the same SED templates 
 \citep{2001ApJ...556..562C} as we used to compute $L_{IR}$ for k-correction to obtain the maximum observable redshift from the $S_{90\mu m}$ flux limit. 
For optical detection limit, we used \citet[][v4\_1]{2003AJ....125.2348B} for k-correction.
 We adoped a redshift where galaxies become fainter than eithter $r$- of  $S_{90\mu m}$-flux limit as $z_{\max}$.

 For each luminosity bin then, the LF is derived as 
\begin{eqnarray}
\phi =\frac{1}{\Delta L}\sum_{i} \frac{1}{V_{\max,i}}w_i, \label{LF}
\end{eqnarray}

\noindent ,where $V_{\max}$  is a comoving volume over which the $i$th galaxy could be observed, and $\Delta L$ is the size of the luminosity bin, 
  and $w_i$ the completeness correction factor of the $i$th galaxy.
 Note that the $V_{\max}$ method considers both the optical and IR flux limits. Therefore, even an IR source with $r>17.7$, a detectable volume of such sources is properly considered by the $V_{\max}$ method as long as IR sources with the same optical-to-IR ratio can be detected at a lower redshift.
 See Appendix of \citet{2001MNRAS.322..262S} and \citet{1980ApJ...235..694A} for more details of a multivariate $V_{\max}$ method.
 A very obscured IR galaxy (with $r>17.7$ even at the lowest redshift of z=0.01), however, would have a detectable volume of zero, and thus, would not be included in the LF computation. The number density of such obscured IR galaxies is a matter of interest itself, but requires deeper optical data. 
 Although we cannot test deeper $r$-band criterion since the SDSS spectroscopic survey stops the current limit, we can test a brighter  $r$-band criterion. We computed the LF using a brighter  $r$-band criterion by one magnitude, at $r<16.7$. The resulting LF was noisier due to the reduced number of galaxies used, but essentially the same as the LF presented in the following sections. The test suggests that the $V_{\max}$ method works well.
 We used completeness correction measured in the FIS bright source catalog release note for $S_{90\mu m}$ flux. This correction is 25\% at maximum, since we only use the sample where the completeness is greater than 80\%. Completeness correction in terms of sky coverage of both surveys is taken into account.

\subsection{Monte Carlo simulation}\label{sec:monte}
 Uncertainties in the LF values stem from various factors such as 
 the finite numbers of sources in each luminosity bin, 
 the k-correction uncertainties,
 and the flux errors. 
 To compute these errors we performed Monte Carlo simulations by creating 150 simulated catalogs, where 
 each catalog contains the same number of sources, but we assign each source new fluxes following a Gaussian distribution centered at fluxes with a width of a measured error.
 Then we measured errors of each bin of the LF based from the variation in the 150 simulations.
 These estimated errors are added in quadrature to the Poisson errors in each LF bin.

\subsection{IR luminosity function}\label{sec:LF}

\begin{figure}
\begin{center}
\includegraphics[scale=0.5]{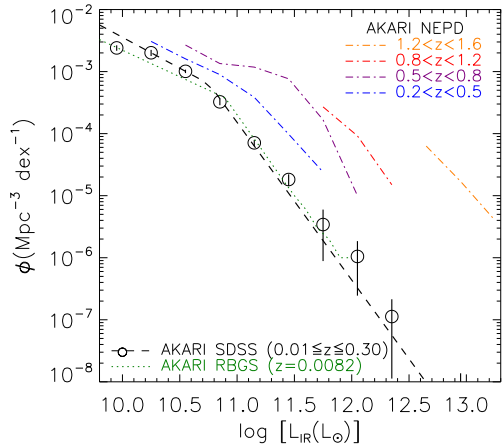}
\end{center}
\caption{
Infrared luminosity function of AKARI-SDSS galaxies. The $L_{IR}$ is measured using the AKARI 9,18,65,90,140 and 160$\mu$m fluxes through an SED fit. Errors are computed using 150 Monte Carlo simulations, added by Poisson error.
The dotted lines show the best-fit double-power law. 
The green dotted lines show IR LF at z=0.0082 by \citet{Goto_RBGS_LF}.
The dashed-dotted lines are higher redshift results from the AKARI NEP deep field \citep{Goto_NEP_LF}.
}\label{fig:LF}
\end{figure}

\begin{figure}
\begin{center}
\includegraphics[scale=0.4]{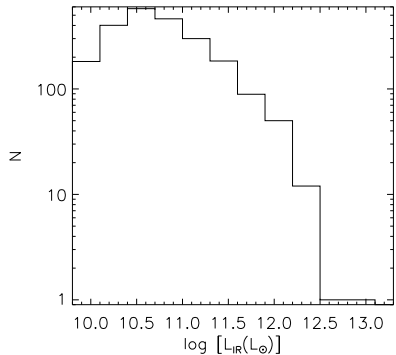}
\end{center}
\caption{An infrared luminosity histogram of galaxies used to compute Fig.\ref{fig:LF}.
}\label{fig:Lhist}
\end{figure}

In Fig.\ref{fig:LF}, we show infrared LF of the AKARI-SDSS galaxies, followed by Fig.\ref{fig:Lhist}, which shows number of galaxies used to compute the LF.
The median redshift of our sample galaxies is z=0.031.
Overplotted in green dotted lines is IR LF of RBGS at z=0.0082 but with re-measured  $L_{IR}$ based on the AKARI 6-band photometry \citep{Goto_RBGS_LF}.
Our LF agrees very well with that from the RBGS \citep{2003AJ....126.1607S,Goto_RBGS_LF}. 
Although it was a concern that the RBGS was $S_{60\mu m}$ selected, the agreement suggests the selection does not affect the total IR LF, perhaps because $60\mu m$ is close to the peak of the dust emission and captures most IR galaxies.
Dotted lines show higher redshift results from the AKARI NEP Deep field  \citep{Goto_NEP_LF}.
Our results (open circles) agree with that from z=0.0082, and show smooth evolution toward higher redshift. 
The faint-end slope of our sample agrees well with immediately higher/lower redshift LFs.

Following \citet{2003AJ....126.1607S}, we fit an analytical function to the LFs.
In the literature, IR LFs were fit better by a double-power law \citep{2006MNRAS.370.1159B,Goto_NEP_LF,Goto_RBGS_LF} or a double-exponential \citep{2004ApJ...609..122P,2006A&A...448..525T,2005ApJ...632..169L} than a Schechter function, which steeply declines at the high luminosity and underestimates the number of bright galaxies.  
In this work, we fit the IR LFs using a double-power law \citep{2006MNRAS.370.1159B} as shown below.
\begin{equation}
 \label{eqn:lumfunc2p}
 \Phi(L)dL/L^{*} = \Phi^{*}\bigg(\frac {L}{L^{*}}\bigg)^{1-\alpha}dL/L^{*}, ~~~ (L<L^{*})
\end{equation}
\begin{equation}
 \label{eqn:lumfunc2p2}
 \Phi(L)dL/L^{*} = \Phi^{*}\bigg(\frac {L}{L^{*}}\bigg)^{1-\beta}dL/L^{*}, ~~~  (L>L^{*})
\end{equation}
\noindent Free parameters are $L^*$ (characteristic luminosity, $L_{\odot}$), $\phi^*$ (normalization, Mpc$^{-3}$), $\alpha$, and $\beta$ (faint and bright end slopes), respectively.
The best-fit values  are summarized in Table \ref{tab:fit_parameters}. 
In Fig.\ref{fig:LF}, the dashed line shows the best-fit double-power law.
The local LF has a break at $L^*$=5.7$\pm0.2\times 10^{10} L_{\odot}$. Understanding how this break ($L^*$) evolves as function of cosmic time, and what causes the break is fundamental to galaxy evolution studies. This work provides an important benchmark in the local Universe.

\begin{table} 
 \centering
  \caption{Best double power-law fit parameters for the AKARI LFs. }\label{tab:fit_parameters}
  \begin{tabular}{@{}cccccccccc@{}}
  \hline
Sample &  $L_{IR}^*$ ($L_{\odot}$)&  $\phi^*(\mathrm{Mpc^{-3} dex^{-1}})$ & $\alpha$ (faint-end)& $\beta$ (bright-end) \\
 \hline
 \hline
Total &  5.7$\pm0.2\times 10^{10}$ &   0.00062$\pm$0.00002   & 1.99$\pm$0.09 &      3.54$\pm$0.09 \\
 \hline
SFG   &   5.0$\pm0.2\times 10^{10}$ &   0.00043$\pm$0.00003   & 1.8$\pm$0.1 &     3.5$\pm$0.3 \\
AGN   &    4.3$\pm0.2\times 10^{10}$ &   0.0004$\pm$0.0001   & 1.4$\pm$0.1 &      3.1$\pm$0.2 \\
\hline
\end{tabular}
\end{table}

\begin{table}
 \centering
  \caption{Local IR luminosity densities by all galaxies, ULIRGs and LIRGs.
}\label{tab:omega}
  \begin{tabular}{@{}lrrlllcccc@{}}
  \hline
  Sample & $\Omega_{IR}^{Total}$  ($L_{\odot}$Mpc$^{-3}$)   &$\Omega_{IR}^{LIRG}(L_{\odot} Mpc^{-3})$ &$\Omega_{IR}^{ULIRG}(L_{\odot} Mpc^{-3})$\\
 \hline
 \hline
Total    &  3.8$^{+5.8}_{-1.2}\times 10^{8}$    & 4.2$^{+0.7}_{-0.8}\times 10^{6}$ &  1.2$^{+0.5}_{-0.4}\times 10^{5}$  \\
SFG    &  5.9$^{+1.4}_{-1.3}\times 10^{7}$  &  2.1$^{+0.7}_{-0.8}\times 10^{6}$  &    1.4$^{+0.2}_{-0.2}\times 10^{5}$ \\
AGN    &  1.9$^{+0.5}_{-0.7}\times 10^{7}$  & 1.9$^{+0.5}_{-0.5}\times 10^{6}$&  2.5$^{+1.3}_{-0.9}\times 10^{5}$	\\
 \hline
\end{tabular}
\end{table}


\subsection{Bolometric IR luminosity density based on the IR LF}\label{sec:omega_IR}
One of the primary purposes in computing IR LFs is to estimate the IR luminosity density,  which in turn is a good estimator of the dust-hidden cosmic star formation density \citep{1998ARA&A..36..189K}, once the AGN contribution is removed. 
 The bolometric IR luminosity of a galaxy is produced by thermal emission of its interstellar matter. In SF galaxies, the UV radiation produced by young stars heats the interstellar dust, and the reprocessed light is emitted in the IR. For this reason, in star-forming galaxies (SFG), the bolometric IR luminosity is a good estimator of the current SFR (star formation rate) of the galaxy.

Once we measured the LF,  we can estimate the total infrared luminosity density by integrating the LF, weighted by the luminosity. We used the best-fit double-power law to integrate outside the luminosity range in which we have data, to obtain estimates of the total infrared luminosity density, $\Omega_{IR}$. Note that outside of the luminosity range we have data ($L_{IR}>10^{12.5}L_{\odot}$ or $L_{IR}<10^{9,8}L_{\odot}$), the LFs are merely an extrapolation and thus uncertain.

The resulting total luminosity density is  $\Omega_{IR}$= (3.8$^{+5.8}_{-1.2})\times 10^{8}$ $L_{\odot}$Mpc$^{-3}$.
Errors are estimated by varying the fit within 1$\sigma$ of uncertainty in LFs.
 Out of  $\Omega_{IR}$, 1.1$\pm0.1$\% is produced by LIRG ($L_{IR}>10^{11}L_{\odot}$), and only 0.03$\pm$0.01\% is by ULIRG ($L_{IR}>10^{12}L_{\odot}$). Although these fractions are larger than z=0.0081, still a small fraction of  $\Omega_{IR}$ is produced by luminous infrared galaxies at z=0.031, in contrast to high-redshift Universe. We will discuss the evolution of  $\Omega_{IR}$ in Section \ref{sec:evolution}.


\subsection{AGN/SFG IR LF}\label{sec:AGN_SFG_IR_LF}

\begin{figure}
\begin{center}
\includegraphics[scale=0.6]{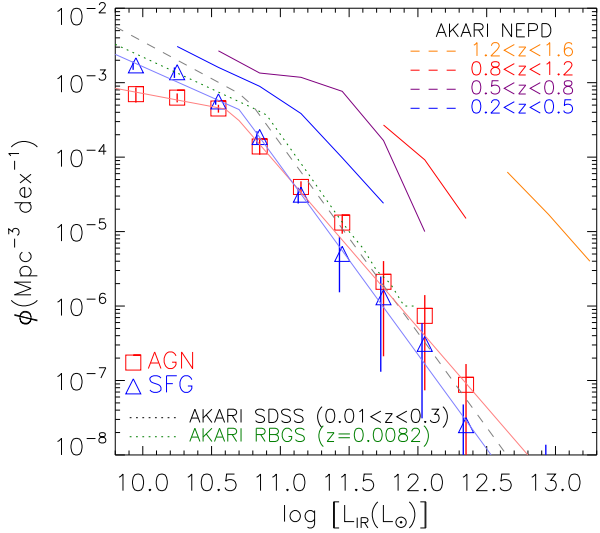}
\end{center}
\caption{
IR LF is separated for star-forming galaxies (blue triangles) and AGN (red squares), using classification in Fig.\ref{fig:BPT}.
Total IR LF in Fig.\ref{fig:LF} is shown with the black dotted lines.
}\label{fig:AGN_LF}
\end{figure}

Once we separated individual galaxies into AGN/SFG, we can construct LFs of each sample.
However, we caution again that our $L_{IR}$ of AGN is total infrared emission of optically-selected AGN, and thus, it includes infrared emission from AGN host galaxies.
In Fig.\ref{fig:AGN_LF}, we show LFs separately for AGN (red squares) and SFG (blue triangles). 
Composite galaxies are included in the AGN LF.
The total IR LF shown in Fig.\ref{fig:LF} is also shown for comparison in the gray dashed line.
It is interesting that both AGN and SFG IR LFs have a break at around log$L_{IR}\sim11.0$.
Then, the IR luminosity density of AGN exceeds that of SFG at log$L_{IR}>11.0$, with flatter slope toward the brigher end.
In contrast, the faint-end slope is steeper for SFG.

 Next, we fit the double-power law (Eqs. \ref{eqn:lumfunc2p} and \ref{eqn:lumfunc2p2}) to the AGN and SF IR LFs, exactly as we did for the total IR LF in Fig.\ref{fig:LF}.
 The best-fit parameters are summarized in Table \ref{tab:fit_parameters}.
 As we mentioned above, the most notable difference is at the bright- and faint-end slopes, where the AGN and SFG LFs dominate, respectively. 
The $L^*$ of AGN is slightly brighter than that of SFG.

%

One of the pioneers in IR AGN LF is \citet{1993ApJS...89..1R}, who selected 893 IR galaxies (of which 118 are AGN) from IRAS 12$\mu$m to minimize wavelength-dependent selection biases towards bluer Seyfert 1 nuclei and redder embedded AGN \citep{1989ApJ...342...83S}. 
 Despite small differences such as selection, AGN classification, and $L_{IR}$ based on the IRAS, their total IR LF of AGN in their Fig.9 is in very good agreement with ours, showing the same AGN dominance at larger $L_{IR}$.

What do these differences in LFs bring to the IR luminosity density, $\Omega_{IR}$, by AGN and SFG?
We estimate  the total infrared luminosity density by integrating the LFs weighted by the luminosity, separately for AGN and SFG. We used the double power law outside the luminosity range in which we have data, to obtain estimates of the total infrared luminosity density, $\Omega_{IR}$, for AGN and SFG.

The resulting total luminosity density ($\Omega_{IR}$) is,
$\Omega_{IR}^{SFG}$=5.9$^{+1.4}_{-1.3}\times 10^{7}$ $L_{\odot}$Mpc$^{-3}$, and
$\Omega_{IR}^{AGN}$=1.9$^{+0.5}_{-0.7}\times 10^{7}$ $L_{\odot}$Mpc$^{-3}$, as summarized in Table \ref{tab:fit_parameters}.
Errors are estimated by varying the fit within 1$\sigma$ of uncertainty in LFs.
The results show that among the total IR luminosity density integrated over all the IR luminosity range,
  75\% ($\frac{\Omega_{IR}^{SFG}}{\Omega_{IR}^{AGN}+\Omega_{IR}^{SFG}}$) of IR luminosity density is emitted by the SFG, and only 
  24\% ($\frac{\Omega_{IR}^{AGN}}{\Omega_{IR}^{AGN}+\Omega_{IR}^{SFG}}$) is by AGN at z=0.031. 
The AGN contribution is larger than z=0.0081, however, is still small compared with higher redshift results \citep{2005ApJ...630...82P,2005ApJ...632..169L,2009A&A...496...57M,Goto_NEP_LF}.

Once we have  $\Omega_{IR}^{SFG}$, we can estimate star formation density emitted in infrared light.
 The SFR and $L_{IR}$ is related by the following equation for a Salpeter IMF, 
$\phi$ (m) 
$\propto m^{-2.35}$ between 
$0.1-100 M_{\odot}$  \citep{1998ARA&A..36..189K}.
\begin{eqnarray}
SFR [M_{\odot} yr^{-1}] =1.72 \times 10^{-10} L_{IR} [L_{\odot}] 
\end{eqnarray}
By using this equation, we obtain SFR density = 1.3$\pm$0.2 $10^{-2}M_{\odot} yr^{-1}$.

If we limit our integration to ULIRG luminosity range  ($L_{IR}>10^{12}L_{\odot}$), we obtain,
    $\Omega_{IR}^{SFG}$(ULIRG)=  1.4$^{+0.2}_{-0.2}\times 10^{5}$ $L_{\odot}$Mpc$^{-3}$, and
    $\Omega_{IR}^{AGN}$(ULIRG)=  2.5$^{+1.3}_{-0.9}\times 10^{5}$ $L_{\odot}$Mpc$^{-3}$.
In other words, at ULIRG luminosity range, AGN explain 
   64\%  ($\frac{\Omega_{IR}^{AGN}(ULIRG)}{\Omega_{IR}^{AGN}(ULIRG)+\Omega_{IR}^{SFG}(ULIRG)}$) of IR luminosity, again showing the AGN dominance at the bright-end.
 
Note that   $\Omega_{IR}^{AGN}$(ULIRG) larger than  $\Omega_{IR}^{Total}$(ULIRG) is apparently inconsistent. This is due to the limitation in extrapolating the double-power law to obtain $\Omega_{IR}$. For example, the AGN LF has a flatter bright-end slope of  beta=3.1$\pm$0.2 than beta of 3.54$\pm$0.09 for total LF. Once extrapolated to larger luminosity at $log L_{IR}>$12.5, the flatter slope could produce larger  $\Omega_{IR}$ once integrated even if all the observed data points for AGN are below those of total IR. To obtain more accurate estimate of $\Omega_{IR}$ we clearly need data points in larger luminosity range than the AKARI observed. In this paper, we tried to obtain best-estimates of each $\Omega_{IR}$ for AGN, SFG, and total. As a result, we caution readers that  $\Omega_{IR}^{Total}$ is not necessarily the algebraical sum of  $\Omega_{IR}^{AGN}$ and $\Omega_{IR}^{SFG}$.

In the LIRG luminosity ($L_{IR}>10^{11}L_{\odot}$), results are
    $\Omega_{IR}^{SFG}$(LIRG)= 2.1$^{+0.7}_{-0.8}\times 10^{6}$ $L_{\odot}$Mpc$^{-3}$, and
    $\Omega_{IR}^{AGN}$(LIRG)= 1.9$^{+0.5}_{-0.5}\times 10^{6}$ $L_{\odot}$Mpc$^{-3}$.
This shows  AGN contribution is already down to 
   47\%  ($\frac{\Omega_{IR}^{AGN}(LIRG)}{\Omega_{IR}^{AGN}(ULIRG)+\Omega_{IR}^{SFG}(LIRG)}$) of IR luminosity in the LIRG range.

\section{Infrared luminosity density and its evolution}
\subsection{Evolution of $\Omega_{IR}^{SFG}$}\label{sec:evolution}

\begin{figure*}
\begin{center}
\includegraphics[scale=1.]{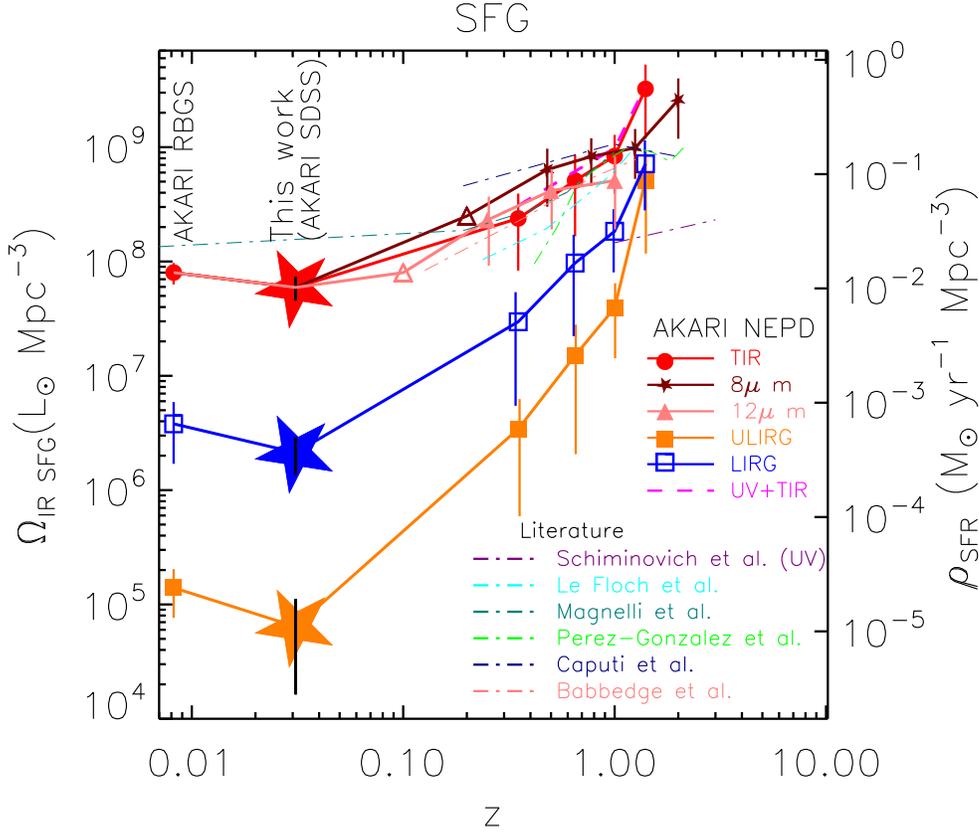}
\end{center}
\caption{
Evolution of IR luminosity density by star-forming galaxies. Results from this work is plotted with stars at z=0.031. 
The red, blue and orange stars show IR luminosity density from all galaxies, from LIRG only, and from ULIRG only. 
Higher redshift results in the solid lines are from the AKARI NEP deep field \citep{Goto_NEP_LF}.
Results at z=0.0082 are from the AKARI RBGS \citep{Goto_RBGS_LF}.
 Shown with different colors are IR luminosity density based on IR LFs (red circles), 8$\mu$m LFs (stars), and 12$\mu$m LFs (filled triangles). The blue open squares and orange filled squares  are for only LIRG and ULIRGs, also based on our $L_{IR}$ LFs.
Overplotted dot-dashed lines are estimates from the literature: \citet{2005ApJ...632..169L}, \citet{2009A&A...496...57M}, \citet{2005ApJ...630...82P}, \citet{2007ApJ...660...97C},   and \citet{2006MNRAS.370.1159B} are in cyan, yellow, green, navy, and pink, respectively.
The purple dash-dotted line shows UV estimate by \citet{2005ApJ...619L..47S}.
The pink dashed line shows the total estimate of IR (IR LF) and UV \citep{2005ApJ...619L..47S}.
}\label{fig:TLD_all}
\end{figure*}

We have separated the  $\Omega_{IR}^{SFG}$ from  $\Omega_{IR}^{AGN}$. 
Now we are ready to examine the evolution of $\Omega_{SFR}$ without contribution from AGN.
In Fig.\ref{fig:TLD_all}, we plot the evolution of  $\Omega_{IR}^{SFR}$ as a function of redshift.
Higher redshift results are taken from \citet{Goto_NEP_LF}, who also tried to exclude AGN using SED fitting.
 Results from the Spitzer survey and GALEX survey are also plotted.
The   $\Omega_{IR}^{SFG}$ shows a strong evolution as a function of redshift. The best-fit linear relation is 
  $\Omega_{IR}^{SFG}\propto$(1+z)$^{4.1\pm0.4}$.
We have removed $L_{IR}$ of optically-classified AGN in this analysis.  A caveat in this process, however, is that $L_{IR}$ of optically-selected AGN perhaps includes some IR emission from star-formation in AGN host galaxies. In this sense, $\Omega_{IR}^{SFG}$ estimated here is a lower-limit.
However, this should not affect $\Omega_{IR}^{SFG}$ more than 30\% since the $\Omega_{IR}^{AGN}$ is smaller than  $\Omega_{IR}^{SFG}$ by a factor of 3 as we show in the next section.

In comparison, our results are in good agreement with previous works shown in the dash-dotted lines in Fig.\ref{fig:TLD_all}.
In the form of $\Omega_{IR}\propto$ (1+z)$^{\gamma}$,
\citet{2005ApJ...632..169L} obtained $\gamma = 3.9\pm 0.4$,
\citet{2005ApJ...630...82P} obtained $\gamma = 4.0\pm 0.2$,
\citet{2006MNRAS.370.1159B} obtained $\gamma = 4.5^{+0.7}_{-0.6}$,
\citet{2009A&A...496...57M} obtained $\gamma = 3.6\pm 0.4$.

Once the IR luminosity density is separated into ULIRG and LIRG contribution,
we found
  $\Omega_{IR}^{SFG} (ULIRG)\propto$(1+z)$^{10.0\pm0.5}$, and
  $\Omega_{IR}^{SFG}(LIRG)\propto$(1+z)$^{6.5\pm0.5}$.
 $\Omega_{IR}^{SFG}(ULIRG)$ shows more rapid evolution than $\Omega_{IR}^{SFG} (LIRG)$, showing importance of luminous IR sources at high redshift.

\subsection{Evolution of $\Omega_{IR}^{AGN}$}\label{sec:agn_evolution}

\begin{figure*}
\begin{center}
\includegraphics[scale=1.]{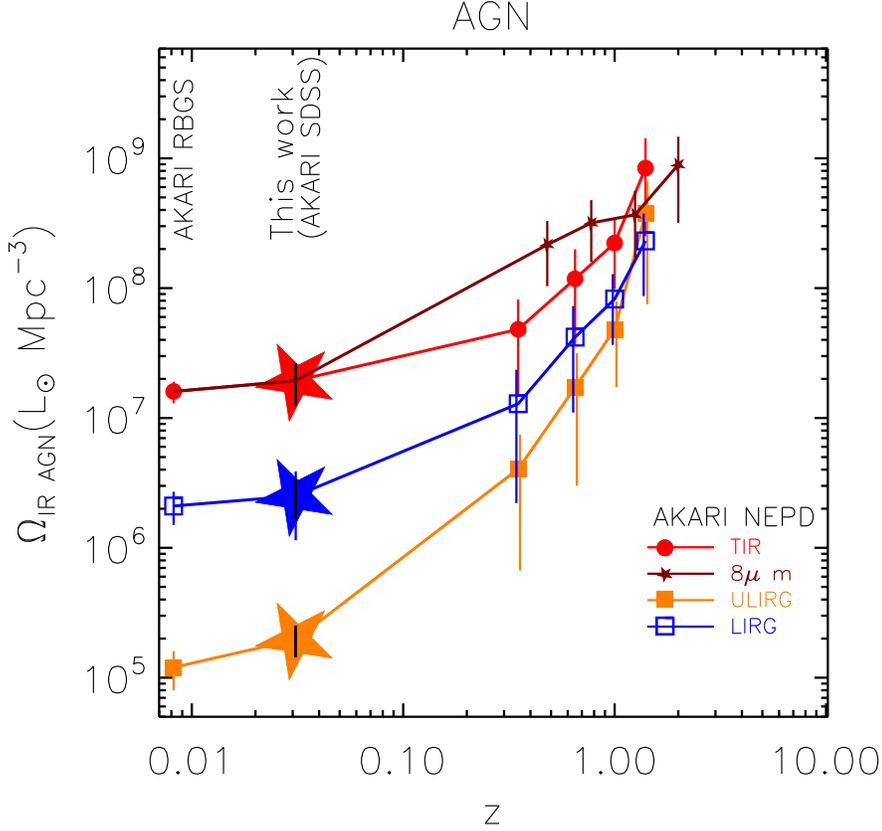}
\end{center}
\caption{
Evolution of IR luminosity density by AGN.
 Results from this work is plotted with stars at the median redshift of z=0.031. 
 The red, blue and orange points show IR luminosity density from all AGN, from LIRG AGN only, and from ULIRG AGN only. 
Higher redshift results are from the AKARI NEP deep field \citep{Goto_NEP_LF}, with contribution from star forming galaxies removed.
 Brown triangles are $\Omega_{IR}^{AGN}$ computed from the 8$\mu$m LFs \citep{Goto_NEP_LF}.
Results at z=0.0082 are from the AKARI RBGS \citep{Goto_RBGS_LF}.
}\label{fig:TLD_AGN_all}
\end{figure*}

In turn, we can also investigate $\Omega_{IR}^{AGN}$. By integrating IR LF$_{AGN}$ in Fig.\ref{fig:AGN_LF}, we obtained 
$\Omega_{IR}^{AGN}$=(1.9$^{+0.5}_{-0.7})\times 10^{7}$ $L_{\odot}$Mpc$^{-3}$.

In Fig.\ref{fig:TLD_AGN_all}, we show the evolution of $\Omega_{IR}^{AGN}$, which shows a strong evolution with increasing redshift. 
 At a first glance, both $\Omega_{IR}^{AGN}$ and $\Omega_{IR}^{SFG}$ show rapid evolution, suggesting that the correlation between star formation and black hole accretion rate continues to hold at higher redshifts, i.e., galaxies and black holes seem to be evolving hand in hand.
 When we fit the evolution with (1+z)$^{\gamma}$, we found 
 $\Omega_{IR}^{AGN}\propto$(1+z)$^{4.1\pm0.5}$.
 A caveat, however, is that $\Omega_{IR}^{AGN}$ estimated in this work likely to include IR emission from host galaxies of AGN, although in optical the AGN component dominates. Therefore, the final conclusion must be drawn from a multi-component fit based on better sampling in FIR by Herschel or SPICA, to separate AGN/SFG contribution to $L_{IR}$.
The contribution by ULIRGs quickly increases toward higher redshift;  By z=1.5, it exceeds that from LIRGs. Indeed, we found 
$\Omega_{IR}^{AGN}(ULIRG)\propto$(1+z)$^{8.7\pm0.6}$ and 
$\Omega_{IR}^{AGN}(LIRG)\propto$(1+z)$^{5.4\pm0.5}$.

\section{Summary}

We have cross-correlated the AKARI IR all sky survey with the SDSS to find 2357 IR galaxies with optical spectra.
Using AKARI's 6-band IR photometry in 9, 18, 65, 90, 140, and 160$\mu m$, we have measured  $L_{IR}$ via SED model fitting. 
The AKARI's 6 bands cover the crucial far-IR wavelengths across the peak of the dust emission, providing us with more accurate $L_{IR}$ measurements than IRAS.
By separating SFG/AGN using optical line ratios, we constructed local IR LFs separately for SFG and AGN.
We also computed local infrared luminosity density through the derived LFs, and compared $\Omega_{IR}^{SFG}$ and $\Omega_{IR}^{AGN}$ to those at higher redshifts.

Our findings are as follows.
\begin{itemize}
 \item The local IR LF with the AKARI data at the median redshift of z=0.031 agrees well with that at z=0.0081, and shows smooth and continuous evolution toward higher redshift results from the AKARI NEP deep field.
 \item By integrating the IR LF weighted by  $L_{IR}$, we obtain  the local cosmic IR luminosity density of $\Omega_{IR}$= (3.8$^{+5.8}_{-1.2})\times 10^{8}$ $L_{\odot}$Mpc$^{-3}$.
 \item The fraction of AGN and composite galaxies show continuous incrase from 25\% to 90\% at $9<log L_{IR}<$12.5.
 \item The SFR based on self-absorption, and extinction corrected $H\alpha$ correlates with $L_{IR}$ for SFGs. 
 \item H$_{\alpha}$/H$_{\beta}$ ratio shows a weak increase with   $L_{IR}$, however with a large scatter.
 \item The $L_{[OIII]}$ correlates well with $L_{IR}$ for AGN.
 \item  The AGN contribution to $\Omega_{IR}$ becomes dominant at $L_{IR}>10^{11}L_{\odot}$, above the break of the both SFG and AGN IR LFs. At $L_{IR}\leq10^{11}L_{\odot}$, SFG dominates IR LFs.
 \item LIRG and ULIRG contribute to $\Omega_{IR}$ a little;
Only  1.1$\pm0.1$\% of $\Omega_{IR}$ is produced by LIRG ($L_{IR}>10^{11}L_{\odot}$), and only 0.03$\pm$0.01\% is by  ULIRG ($L_{IR}>10^{12}L_{\odot}$) in the local Universe. 
 \item  Compared with high redshift results from the AKARI NEP deep survey, we observed a strong evolution of 
$\Omega_{IR}^{SFG}\propto$(1+z)$^{4.1\pm0.4}$, and  $\Omega_{IR}^{AGN}$ $\propto$(1+z)$^{4.1\pm0.5}$. 
\end{itemize}

\section*{Acknowledgments}

We thank the anonymous referee for many insightful comments, which significantly improved the paper.


T.G. acknowledges financial support from the Japan Society for the Promotion of Science (JSPS) through JSPS Research Fellowships for Young Scientists.



%

This research is based on the observations with AKARI, a JAXA project with the participation of ESA.

The authors wish to recognize and acknowledge the very significant cultural role and reverence that the summit of Mauna Kea has always had within the indigenous Hawaiian community.  We are most fortunate to have the opportunity to conduct observations from this sacred mountain.

TTT has been supported by Program for Improvement of Research
Environment for Young Researchers from Special Coordination Funds for
Promoting Science and Technology, and the Grant-in-Aid for the Scientific
Research Fund (20740105) commissioned by the Ministry of Education,
Culture,
Sports, Science and Technology (MEXT) of Japan.
TTT has been also partially supported from the Grand-in-Aid for the Global
COE Program ``Quest for Fundamental Principles in the Universe: from
Particles to the Solar System and the Cosmos'' from the MEXT.

%
%





\label{lastpage}

\end{document}